\documentstyle[aps]{revtex}
\begin{document}
\newcommand{\beq}{\begin{equation}}
\newcommand{\eeq}{\end{equation}}
\newcommand{\beqn}{\begin{eqnarray}}
\newcommand{\eeqn}{\end{eqnarray}}
\newcommand{\bmath}{\begin{mathletters}}
\newcommand{\emath}{\end{mathletters}}
\twocolumn[\hsize\textwidth\columnwidth\hsize\csname @twocolumnfalse\endcsname 
\title{Hole Superconductivity in $Mg B_2$: a high $T_c$ cuprate without $Cu$}
\author{J. E. Hirsch }
\address{Department of Physics, University of California, San Diego\\
La Jolla, CA 92093-0319}

\date{\today} 
\maketitle 
\begin{abstract} 
The theory of hole superconductivity explains high temperature superconductivity in
cuprates as driven by pairing of hole carriers in oxygen $p\pi$ orbitals in 
the highly negatively charged $Cu-O$ planes. The pairing mechanism is hole undressing 
and is Coulomb-interaction driven. We propose that the planes of $B$ atoms in 
$Mg B_2$ are akin to the $Cu-O$ planes without $Cu$, and that the recently observed 
high temperature superconductivity in $Mg B_2$ arises similarly from undressing 
of hole carriers in the planar boron $p_{x,y}$ orbitals in the negatively charged
$B^-$ planes. Doping $Mg B_2$ with electrons and with
holes should mirror the behavior of underdoped and overdoped high $T_c$
cuprates respectively. We discuss possible ways to achieve higher transition temperatures
in boron compounds based on this theory.
\end{abstract}

\vskip2pc]
 
\section{Introduction}

Superconductivity with a remarkably high transition temperature ($\sim 40K$) has
recently been discovered in $Mg B_2$\cite{super}. It has been proposed
that it results from strong electron-phonon interaction and high phonon
frequency due to the light ionic masses\cite{elph}, and indeed measurement
of an isotope effect has been reported\cite{isotope}. Here we propose instead
that the phenomenon is another clear example of the mechanism of hole
superconductivity at work\cite{holemech}. The theory of hole superconductivity
proposes that this is the $universal$ mechanism of superconductivity
in solids\cite{holeth}.

No measurement of the Hall coefficient of $MgB_2$ has yet been reported in
the literature to our knowledge. The Hall coefficient for other metal diborides
with the same crystal structure ($YB_2$, $TiB_2$, $VB_2$, $ZrB_2$, $NbB_2$, etc.)
has been measured and found to be negative\cite{hall1,hall2}. Correspondingly
we expect no superconductivity in those compounds (no superconductivity
in those compounds has ever been detected to our knowledge). Instead, we
expect the Hall coefficient of $MgB_2$, certainly for transport in the planes,
to be positive, indicating the dominant role of hole carriers in the normal 
state transport expected for  superconductors within our theory.
In other words, we expect the family of metal diboride compounds to be another clear
example of the Chapnik-Feynman empirical rule\cite{chapnik,feynman,correl} that 
superconductivity exists when the
normal state transport is hole-like and does not exist when it is electron-like.

In the theory under consideration here\cite{holemech,holeth}, superconductivity is
an 'undressing' transition, and it can only be driven by carriers in
bands that are almost full.  It is argued 
that the dressing of quasiparticles in an electronic energy band, due to 
electron-electron interactions, is an increasing function of the
electronic band occupation. When the Fermi level is near the top of
the band, the carriers (hole carriers) are most heavily dressed, and the normal state
transport is largely incoherent; when these heavily
dressed hole carriers pair they partially undress, and this is the driving
force for superconductivity. The superconducting condensation energy is
kinetic in origin, as undressed carriers have a lower effective mass.
Similarly, undressing also occurs upon 
hole doping a nearly full band in the normal state. If the Fermi level is
not close to the top of the band, carriers are not as heavily dressed in the
normal state and do not gain enough by pairing to overcome Coulomb repulsion,
and superconductivity disappears.

\section{What makes $Mg B_2$ a high temperature superconductor}

The crystal structure of $MgB_2$ is the so-called $Al B_2$ structure: 
honeycomb layers of boron atoms alternate with hexagonal layers of $Mg$ atoms.
B atoms in different planes are on top of each other, and the $Mg$ atoms are
at the center of the hexagons defined by the $B$ atoms. It was proposed
long ago that in metal diborides the boron atoms accept electrons from
the metal, and the boron planes become negatively charged\cite{lipscomb}.
The simplest ionic picture would suggest that in $MgB_2$  the $Mg$ donates
two electrons to the $B$ planes, and the ionic compound $Mg^{++}(B^-)_2$
results, which will be  metallic due to boron band overlap.
This picture is supported by band structure calculations.

Within the theory of hole superconductivity what drives superconductivity in
the cuprates is hole transport through  the planar oxygen $p\pi$ orbitals\cite{tang}.
That band is full in the undoped cuprates, and we have proposed that when the
materials are hole doped, hole carriers will go predominantly into that band
(although transport through the O $p\sigma$-Cu $d_{x^2-y^2}$ orbitals may
also occur\cite{twoband}). For the electron-doped cuprates, we have proposed that
holes in the planar O $p\pi$ orbitals are induced by electron doping\cite{electron1}
and that they become mobile when oxygen is removed through annealing in a 
reducing atmosphere\cite{electron2}.

Band structure calculations for the doped cuprates are complicated due to the strong
Coulomb repulsion on the $Cu$ atoms, and so far they have $not$ shown evidence
that the Fermi level cuts an oxygen $p\pi$ band near its top. In contrast,
band structure calculations for $MgB_2$ are simpler and expected to be accurate,
and several calculations have been reported in the literature, using
different methodologies, that are in essential agreement\cite{band1,band2,band3,elph}.

Figs. 1 and 2 show band structure calculation results of Armstrong et al 
for $MgB_2$\cite{band1}
and $AlB_2$\cite{band5} respectively. The essential fact that makes $MgB_2$ a superconductor,
according to our theory, is that the Fermi level at the $\Gamma$ point cuts two
bands right near the top, as seen in Fig. 1 (bands 3 and 4 in the terminology of
Armstrong et al\cite{band1}). These two bands give rise to nearly cylindrical
hole-like Fermi surfaces around the $\Gamma$ point (see also Fig. 3 of 
Ref. 2), indicating that the transport from these hole 
carriers will be dominantly
in-plane. The atomic states giving rise to these bands are dominantly boron
planar  $p_{x,y}$ orbitals\cite{isotope,band1}. Within the theory of hole superconductivity,
it is those heavily dressed hole $p_{x,y}$ carriers that 
give rise to the high temperature superconductivity of $MgB_2$. As emphasized by Armstrong et
al, the main difference between $MgB_2$ and $AlB_2$ is that in the latter compound
these $p_{x,y}$ orbitals are completely filled and do not contribute to conduction.
The conduction in $AlB_2$ is mainly electron-like, as suggested by Fig. 2 and by the
measured negative Hall coefficient of the similar compound $YB_2$\cite{hall1}.
Hence $AlB_2$ is not a superconductor.

The existence of a hole-like Fermi surface is a necessary condition for
superconductivity within our model. However to obtain high temperature superconductivity
the presence of negatively charged conducting structures is also found to be favorable,
as negative ions give rise to larger values of the
parameter $\Delta t$ that drives superconductivity\cite{porb}. Thus the fact that the
boron planes in $MgB_2$ are essentially $B^-$, i.e. are highly negatively
charged, naturally fits within this picture.
 The same situation is found in the high $T_c$ oxides, where the
$Cu-O$ planes are also highly negatively charged, with  two extra negative charges 
per unit cell ($Cu^{++}(O^{=})_2$).

Figure 3 shows the generic behavior of $T_c$ versus hole concentration 
predicted by our theory\cite{holemech,electron1}.
The magnitude of the parameter $\Delta t$ used, that drives superconductivity
in our model, is $\Delta t \sim 0.37 eV$ for the case of Fig. 3, well within 
the range obtained from first-principles calculations of this parameter
for p-orbitals and negatively charged ions\cite{porb}.
According to the Mulliken population analysis of Armstrong and Perkins\cite{band1}
the hole occupation in the boron $p_{x,y}$ orbitals in $MgB_2$ is $0.07$. Although
the position of the maximum $T_c$ in our model can vary somewhat with parameters,
it generally occurs for lower hole concentrations, as seen in the
example in Fig. 3. This would imply that $MgB_2$ is slightly overdoped, so that
a small decrease in the hole carrier concentration would increase $T_c$
up to a maximum of around $50K$. Doping
with electrons, for example in a compound $Mg_{1-x}Al_xB_2$ (assuming it
forms with the same structure) would bring $T_c$ over the maximum and drive
it to zero in the underdoped regime. Doping with holes, for example with
$Mg_{1-y}Na_yB_2$ (assuming it forms with the same structure) would drive
$T_c$ to zero in the overdoped regime.

The theory of hole superconductivity predicts a cross-over from strong to
weak coupling regimes as the hole concentration increases\cite{strong},
a scenario which is in qualitative agreement with observations in the
cuprates. We may expect to see a similar scenario (although less
pronounced) in the doped magnesium diborides. In the underdoped regime
($Mg_{1-x}Al_xB_2$) increasingly incoherent transport and higher resistivity,
decreasing coherence length and increasing gap ratio as $x$ increases,
possibly even pseudogap behavior and charge inhomogeneity. This should
only occur for a small range of $x$ however, after which the $p_{x,y}$ bands
will become full and the behavior will change sharply for larger $x$:
superconductivity will dissappear, the Hall coefficient will become negative,
the transport will become coherent, and the resistivity will decrease as
$x$ increases further with increasing electron carriers. The lattice stability should
also increase in this regime of large $x$. In contrast, in the overdoped
regime ($Mg_{1-y}Na_yB_2$) we expect increasingly coherent behavior as
$y$ increases, with lower resistivity, increasing superconducting
coherence length diverging
as $T_c$ approaches zero, and gap ratio close to the BCS weak coupling value.
The London penetration depth should decrease monotonically as the hole
concentration increases from underdoped to overdoped\cite{london}.

\section{Effect of pressure}

Within our theory a decreasing $B-B$ intraplane distance should increase $T_c$
\cite{holemech}, as seen in Fig. 3. Similarly we have argued that in
high $T_c$ oxides the intrinsic effect of increase of $T_c$ with pressure is
caused by decreasing intraplane $O-O$ distance\cite{holepressure}. This is because superconductivity is
driven by the correlated hopping parameter $\Delta t$ that depends
exponentially on  interatomic distance. Hence we expect application of pressure
in the plane direction in $MgB_2$ will increase $T_c$. Hydrostatic pressure
should also increase $T_c$ assuming it leads to a decrease in the
$B-B$ intraplane distance. However the situation could be more complicated if
charge transfer between the $B$ planes and the metal occurs under pressure, in 
which case the sign and magnitude of the change in $T_c$ would depend on the 
sign of the charge transfer, whether the system is in the overdoped or 
underdoped regime, and on the relative weight of the change induced by
charge transfer and change induced by changing interatomic distances.
All of these effects could be disentangled by measuring changes in
lattice constants and Hall coefficient under pressure.

\section{How to achieve higher $T_c$'s in the diborides}

Various calculations of band structures in diborides suggest that
a rigid band picture works reasonably well\cite{band1,band2,band3,band5,band4}.
In the transition metal diborides it is found that the Fermi level states 
are dominantly of metal $3d$ character\cite{band4}, in contrast to the 
main group diborides where the boron $2p_{x,y}$ orbitals dominate the
Fermi level states. According to the theory discussed here, the latter
situation is the favorable one for superconductivity, and the position of
the Fermi level in $MgB_2$ is close to optimal. Hence what remains to
be optimized is the interatomic distances, which may be achieved
by 'chemical pressure'.

Consider the compound $BeB_2$\cite{beb2,sands}. Its structure is similar although
not identical to that of $MgB_2$, however the interatomic distances
should be significantly smaller than those in $MgB_2$ due to the smaller size of
the $Be$ atom. The lattice constants for $BeB_2$ obtained by averaging
experimental data over a larger unit cell have been inferred to be
$a=2.94A$, $c=2.87A$\cite{beb2p}, while those of $MgB_2$ are 
$a=3.084A$, $c=3.522A$\cite{isotope}. Hence the crucial $B-B$ intra-plane
distance should be about $5\%$ shorter in $BeB_2$, allowing in principle
for considerably higher $T_c$.

However even though $Be$ and $Mg$ are both nominally divalent the ionization
potentials of $Be$ are significantly larger: the first and second ionization 
potentials are $E_I=9.32eV$, $E_{II}=18.21eV$ for $Be$, and 
$E_I=7.64eV$, $E_{II}=15.03eV$ for $Mg$\cite{kittel}. Hence the charge
transfer from $Be$ to $B$ in $BeB_2$ will be less than that from $Mg$ to $B$
in $MgB_2$, and
the Fermi level in $BeB_2$ will be lower than shown in Fig. 1, probably
beyond the regime where superconductivity occurs. Hence the system
$BeB_2$ needs to be doped with electrons, and we suggest to achieve that
through the compound $Be_{1-x}Al_xB_2$. If this compound forms in
the $AlB_2$ structure, we predict that with increasing $x$ the $T_c$ versus
$n_h$ curve of Fig. 3 will be mapped from overdoped to underdoped, with a
maximum $T_c$ significantly larger than in the $MgB_2$ structure.

Similarly a $LiB_2$ or $NaB_2$ compound with the same structure and doped with
electrons, e.g. $Li_{1-x}Al_xB_2$, would map the $T_c$ versus $n_h$ curve,
for larger values of $x$ ($x>1/2$). The $Li$ case should yield higher $T_c's$
than the $Na$ case due to the smaller interatomic distances.

Finally we comment on the transition metal diborides. Several compounds of the
form $MB_2$ form with the $AlB_2$ structure, e.g. 
$M=Sc, Ti, V, Cr, Y, Zr, Nb, Mo, Hf, Ta, W$\cite{hall1,hall2}. All are reported to have
negative Hall coefficient\cite{hall1,hall2} and the transport is dominated by 
carriers in the metal $d$ orbitals\cite{band3,band4}. The band structure calculations of 
Armstrong\cite{band4} show that the Fermi level is above the $\Gamma_5$ point 
in Figure 1, and approaches that point as one moves to the right and down
in the periodic table. Doping these materials with holes, for example by forming
intermetallic compounds where transition metals are partially 
substituted by alkali or alkali earths, should bring the Fermi level
down to the $B$ $p_{x,y}$ bands and give rise to superconductivity.
However these cases should be like the two-band situation of ref.\cite{twoband},
where the hole carriers drive the system superconducting in the
presence of a large number of electron carriers, and $T_c$'s should be
considerably lower than in $MgB_2$.

\section{Isotope effect}
The presence of an isotope effect is usually considered evidence that 
electron-phonon coupling drives superconductivity. However, an isotope effect
is also expected in the mechanism considered here. The parameter that
drives superconductivity, $\Delta t$, depends sensitively on interatomic
distances, as does the single particle hopping $t$\cite{holeth}. Let us assume we have
\bmath
\beq
t(q)=t+\gamma q
\eeq
\beq
\Delta t(q)=\Delta t +\gamma ' q
\eeq
\emath
where $q$ is an optic phonon mode, and $\gamma '=\gamma \Delta t/t$. The zero-point
motion of the atoms will yield a small increase in the effective
$\Delta t$, that depends on the ionic mass:
\beq
\Delta t_{eff}=\sqrt{<\Delta t>^2}=\Delta t+\frac{\gamma'^2}{\Delta t}<q^2>=
\Delta t+\gamma^2\frac{\Delta t}{t^2}<q^2>
\eeq
Using $<q^2>=\hbar\omega /2K$, with $\omega$ the phonon frequency and 
$K$ the force constant, we find that if the ionic mass $M$ changes by
$\delta M$, the parameter $\Delta t_{eff}$ changes by
\beq
\delta (\Delta t_{eff})=\frac{\alpha^2}{4K}\frac{\Delta t}{t^2}\hbar\omega
\frac{\delta M}{M}
\eeq
Finally, with $D=2zt$ the bandwidth ($z$ the number of nearest neighbors)
and $\lambda=\gamma^2/KD$ the dimensionless electron-phonon coupling we
have
\beq
\delta (\Delta t_{eff})=z\lambda \frac{\Delta t}{t}\frac{\hbar \omega}{2}\frac{\delta M}{M}
\eeq
and the change in $T_c$ is given by
\beq
\frac{\partial ln T_c}{\partial ln M}=\frac{\partial ln T_c}{\partial ln \Delta t}\times
\frac{z\lambda \hbar \omega}{2t}
\eeq
For the parameters used in Figure 3, $\partial ln T_c/\partial ln \Delta t=14.5$,
$t\sim 0.63 eV$, and using $\omega =750K$ as an estimate of the
phonon frequency\cite{isotope} we find for the isotope coefficient
\beq
\alpha \equiv \frac {\partial ln T_c}{\partial ln M} \sim 2.2 \lambda
\eeq
so that to reproduce the measured isotope shift $\alpha _B\sim 0.26$\cite{isotope}
would require a rather small $\lambda$, $\lambda \sim 0.12$.
The estimated value of $\lambda$ in ref. \cite{elph} is much larger,
$\lambda \sim 0.7$.

While the calculation discussed here is not expected to be very accurate,
it illustrates the fact that the mechanism of hole superconductivity is
very sensitive to ionic vibrations and generically leads to a positive
isotope effect due to the increased effective $\Delta t$ from larger zero-point 
vibrations. Eq. (6) and the measured isotope shift suggest that the actual
$\lambda$ in the material may be substantially smaller than estimated in
ref. \cite{elph}. Alternatively, the isotope shift estimate Eq. (6) may be
reduced by the fact that a smaller ionic mass would also be expected to
lead to a slightly larger average interatomic distance, an effect
which would partially compensate the increase in effective $\Delta t$ 
estimated here.

\section{Conclusions}

We have proposed here an interpretation of the observed high temperature
superconductivity in $MgB_2$ based on the theory of hole superconductivity.
According to this theory, the essence of high $T_c$ cuprates is metallic
oxygen, more specifically holes conducting through nearly full oxygen
$p\pi$ planar orbitals in the highly negatively charged $Cu-O$ planes\cite{tang}. 
Similarly, holes in nearly filled boron  planar $p_{x,y}$ orbitals in the negatively 
charged $B^-$ planes should drive superconductivity in $MgB_2$. 
In this sense, $MgB_2$ is then a beautiful realization of the essential
physics of superconductivity in cuprates, without the complications
of $Cu$ $d_{x^2-y^2}$ orbitals, antiferromagnetism, chains, etc., that
we have argued obscure the physics and are non-essential in the
cuprates\cite{holemech,holeth}. Already in our early work we noted in connection
with high $T_c$ cuprates 
``a moment on the cation (like $Cu^{2+}$) is not needed, in particular a 
closed shell should do'', and ``The high $T_c$ oxide structures with
$Mg$ in place of $Cu$, if they formed, would be excellent candidates to
support our picture and rule out other mechanisms''\cite{hole1}.
 Remarkably, the band structure calculations in the borides and their 
interpretation\cite{elph,band1} directly support the scenario expected in our 
theory to be favorable for hole superconductivity.

We expect that the various effects predicted by the theory of hole
superconductivity for the cuprates\cite{holemech,holeth} should exist, and be of appreciable
magnitude in $MgB_2$, due to its high $T_c$, in particular:
(1) tunneling asymmetry of universal sign, i.e. larger current for
negatively biased sample; (2) charge imbalance of quasiparticle
excitations (quasiparticles have a net positive charge), resulting in
particular in positive thermopower for NIS and SIS tunnel junctions;
(3) apparent violation of low energy conductivity sum rule, i.e. a 
larger $\delta$-function weight (smaller London penetration depth)
than expected from the missing area in the low frequency
conductivity; (4) color change, i.e. transfer of optical spectral weight
from high frequencies, up to the visible range, down to low frequencies
upon onset of superconductivity; (5) evidence for 'undressing' in
angle-resolved photoemission\cite{ding}, i.e. emergence of a sharp quasiparticle
peak in the superconducting state from a weaker peak and an incoherent background
in the normal state, as well as evidence of increased coherence upon
hole doping in the normal state. 

Furthermore, as discussed in previous sections, we expect $T_c$ to be an increasing
function of pressure applied in the plane direction, positive Hall
coefficient for $MgB_2$, and $T_c$ versus hole concentration of the generic
form given by Figure 3, with strong coupling to weak coupling crossover
in various properties as function of increasing hole doping in the 
various diboride compounds discussed in the text. As we noted in
earlier work\cite{electron1,carrier2}, the same generic $T_c$ versus hole
concentration behavior is seen in transition metal alloys, a 
phenomenon known as ``Matthias' rules''\cite{matthias}.

Of course there will be other theories proposed to explain the superconductivity
of $MgB_2$ and related compounds, and in particular an electron-phonon
one has already been advanced\cite{elph}. Even though an isotope effect
may appear to favor such a theory, we have pointed out here that
such an effect is also expected within the theory of hole superconductivity.
Fortunately  all theories should not
yield identical predictions as to which other diborides should give rise
to high temperature superconductivity, nor about what their properties are. 
Spelling out such predictions
clearly, preferably before the experiments are performed, should make it
easier to ascertain the relative merits of the various theories. 

In closing we remark that the theory of superconductivity through hole 
undressing discussed here predicts the symmetry of the superconducting state for all
superconductors to be s-wave. We suggest that the apparent evidence for
d-wave superconductivity in the cuprates may be connected to the
presence of Cu d orbitals near the Fermi level. Since no such orbitals
exist in $MgB_2$ we expect that clear evidence for s-wave superconductivity
will be experimentally found.

\begin{figure}
\caption { Results of band structure calculation for $MgB_2$ in the plane directions
reproduced from Ref. 16 by Armstrong et al. Note the hole pockets at the
$\Gamma_5$ point (where the arrow points). 
They give rise to cylindrical hole Fermi surfaces  in the
c direction for the third and fourth bands.
}
\label{Fig. 1}
\end{figure}

\begin{figure}
\caption { Band structure calculation results  for $AlB_2$ in the plane directions
reproduced from Ref. 19 by Armstrong et al. In contrast to Figure 1, the hole
pockets at the $\Gamma $ point no longer exist as the $\Gamma_5$ point is
considerably below the Fermi level. The Fermi surface for this case is
predominantly electron-like.
}
\label{Fig. 2}
\end{figure}

\begin{figure}
\caption { $T_c$ versus hole concentration $n_h$ in the model of hole superconductivity
for a two-dimensional case. This behavior is generic for this model.
Values for the bandwidth, correlated hopping parameter,
on-site and nearest neighbor repulsion used were $D=5eV$, $\Delta t=0.3725eV$, $U=5eV$, $V=0$
respectively. The dashed line indicates the behavior expected under application
of physical or chemical pressure, with the parameter $\Delta t$ increased to $0.375eV$.
}
\label{Fig. 3}
\end{figure}

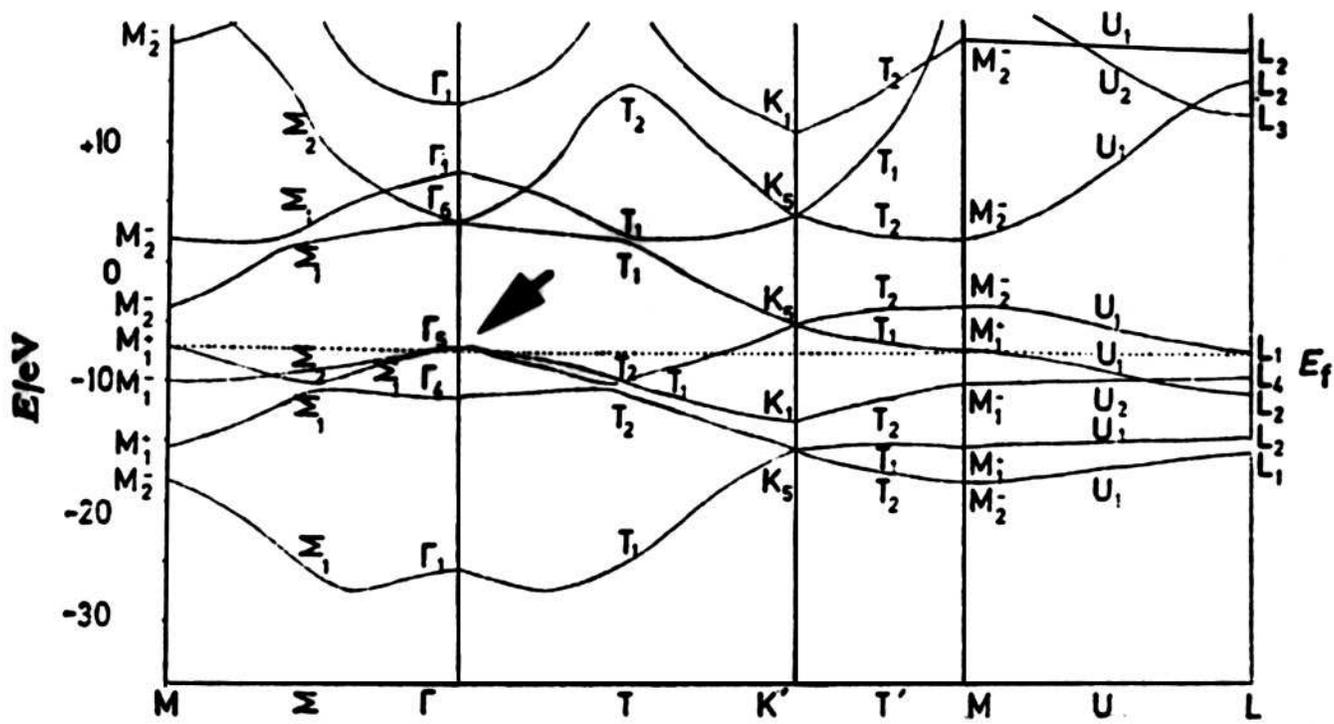
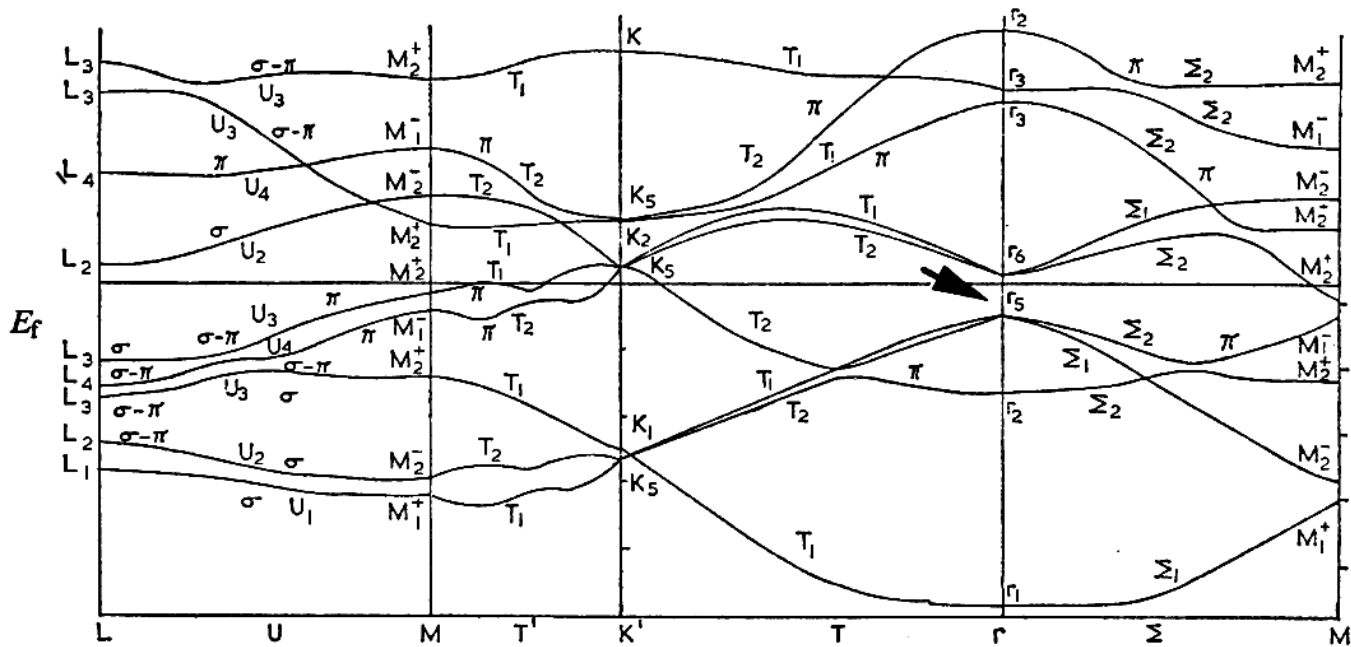
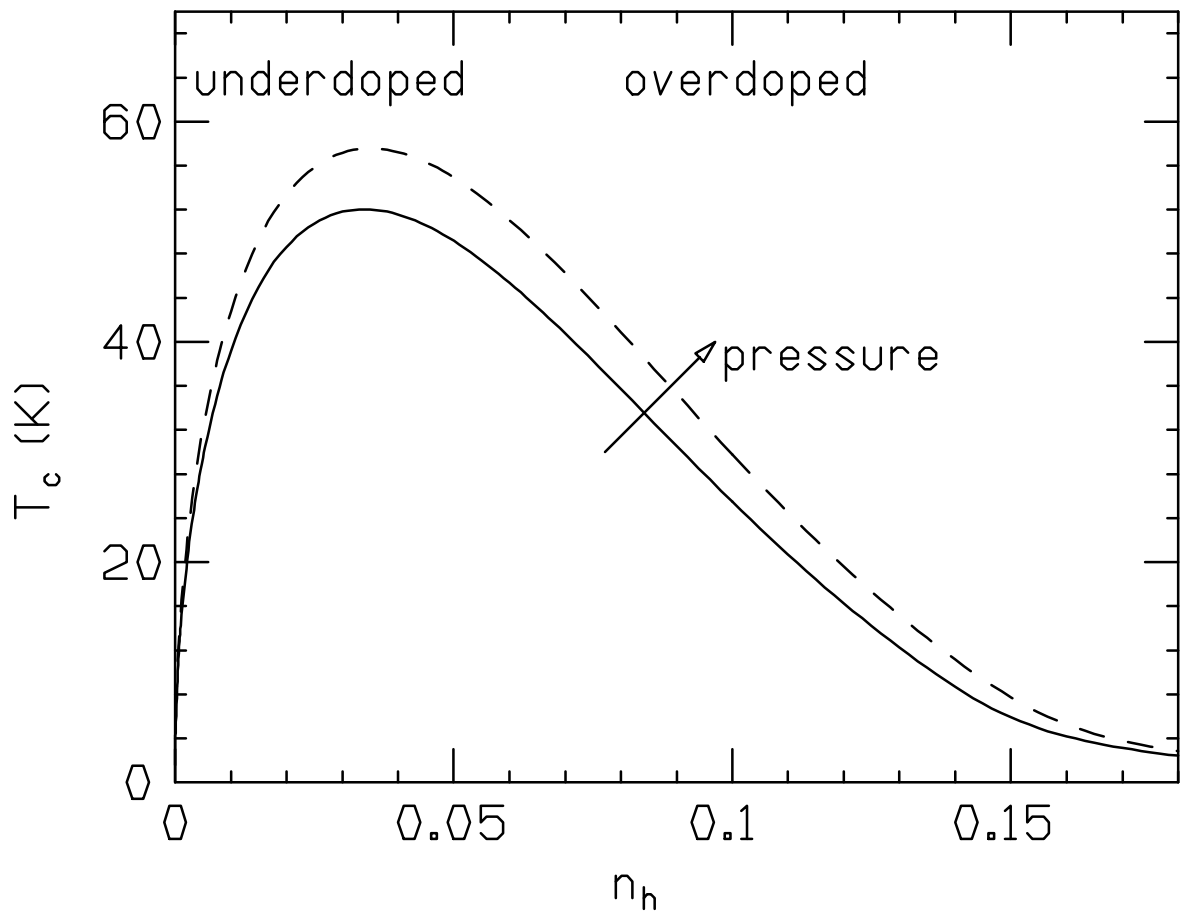
\end{document}